\documentclass[aps,twocolumn,showpacs,amsmath,floatfix]{revtex4}
\usepackage{graphicx}%
\usepackage{dcolumn}
\usepackage{color}
\usepackage{amsmath}
\usepackage{here}

\begin{document}
\begin{sloppy}
\newcommand{\be}{\begin{equation}}
\newcommand{\ee}{\end{equation}}
\newcommand\bibtranslation[1]{English translation: {#1}}
\newcommand\bibfollowup[1]{{#1}}

\newcommand\pictc[5]{\begin{figure}
                       \centerline{
                       \includegraphics[width=#1\columnwidth]{#3}}
                   \protect\caption{\protect\label{fig:#4} #5}
                    \end{figure}            }
\newcommand\pict[4][.8]{\pictc{#1}{!tb}{#2}{#3}{#4}}
\newcommand\rpict[1]{\ref{fig:#1}}

\newcommand\leqt[1]{\protect\label{eq:#1}}
\newcommand\reqtn[1]{\ref{eq:#1}}
\newcommand\reqt[1]{(\reqtn{#1})}

\newcounter{Fig}
\newcommand\pictFig[1]{\pagebreak \centerline{
                   \includegraphics[width=\columnwidth]{#1}}
                   \vspace*{2cm}
                   \centerline{Fig. \protect\addtocounter{Fig}{1}\theFig.}}

\title{Birefringent left-handed metamaterials and perfect lenses}

\author{Alexander A. Zharov$^{1,2}$}
\author{Nina A. Zharova$^{1,3}$}
\author{Roman E. Noskov$^2$}
\author{Ilya V. Shadrivov$^1$}
\author{Yuri S. Kivshar$^1$}

\affiliation{$^1$Nonlinear Physics Centre, Research School of
Physical Sciences and Engineering, Australian National University,
Canberra ACT 0200, Australia \\
$^2$Institute for Physics of Microstructures, Russian Academy of
Sciences, Nizhny Novgorod 603950, Russia\\
$^3$Institute of Applied Physics, Russian Academy of Sciences,
Nizhny Novgorod 603600, Russia}

\begin{abstract}
We describe the properties of birefringent left-handed
metamaterials and introduce the concept of {\em a birefringent
perfect lens}. We demonstrate that, in a sharp contrast to the
conventional left-handed perfect lens at $\epsilon=\mu=-1$, where
$\epsilon$ is the dielectric constant and $\mu$ is the magnetic
permeability, the birefringent left-handed lens can focus either
TE or TM polarized waves or both of them, allowing a spatial
separation of the TE and TM images. We discuss several
applications of the birefringent left-handed lenses such as the
beam splitting and near-field diagnostics at the sub-wavelength
scale.
\end{abstract}


\maketitle

A great interest to the subwavelength imaging is explained by a
number of potential applications, including lithography and data
storage, which could use the resolution abilities better than the
wavelength. One of the potential candidates for improving the
resolution of an imaging system is the so-called ``perfect
lens''~\cite{ref1_anis} based on the concept of the left-handed
metamaterials~\cite{ref3_anis,review}. The possibility of a
perfect lens whose resolution is not limited by the classical
diffraction limit has been a subject of intense debates by the
scientific community during the past three years, and at first met
with considerable opposition~\cite{critics}. However, many
difficulties raised by the critics have been answered by
clarification of the concept and its limitations~\cite{more}, by
numerical simulations~\cite{numerics}, and in recent experiments,
e.g. with a negative-index material assembled from discrete
elements arranged on a planar circuit board~\cite{george}.

The ``perfect lens'' is created by a slab of left-handed
metamaterial with $\epsilon= \mu =-1$, where $\epsilon$ is the
dielectric constant and $\mu$ is the magnetic permeability.
Veselago predicted~\cite{ref3_anis} that such a material would
have a negative refractive index of $n=-\sqrt{\epsilon \mu}=-1$,
and a slab of such a material would act as a lens refocusing {\em
all rays} from a point source on one side of the slab into a point
on the other side of the slab (see Fig.~\rpict{fig1}). Later,
Pendry has shown \cite{ref1_anis} that such a lens can reconstruct the near field of the source, and as a result it can create an ideal image.

\pict{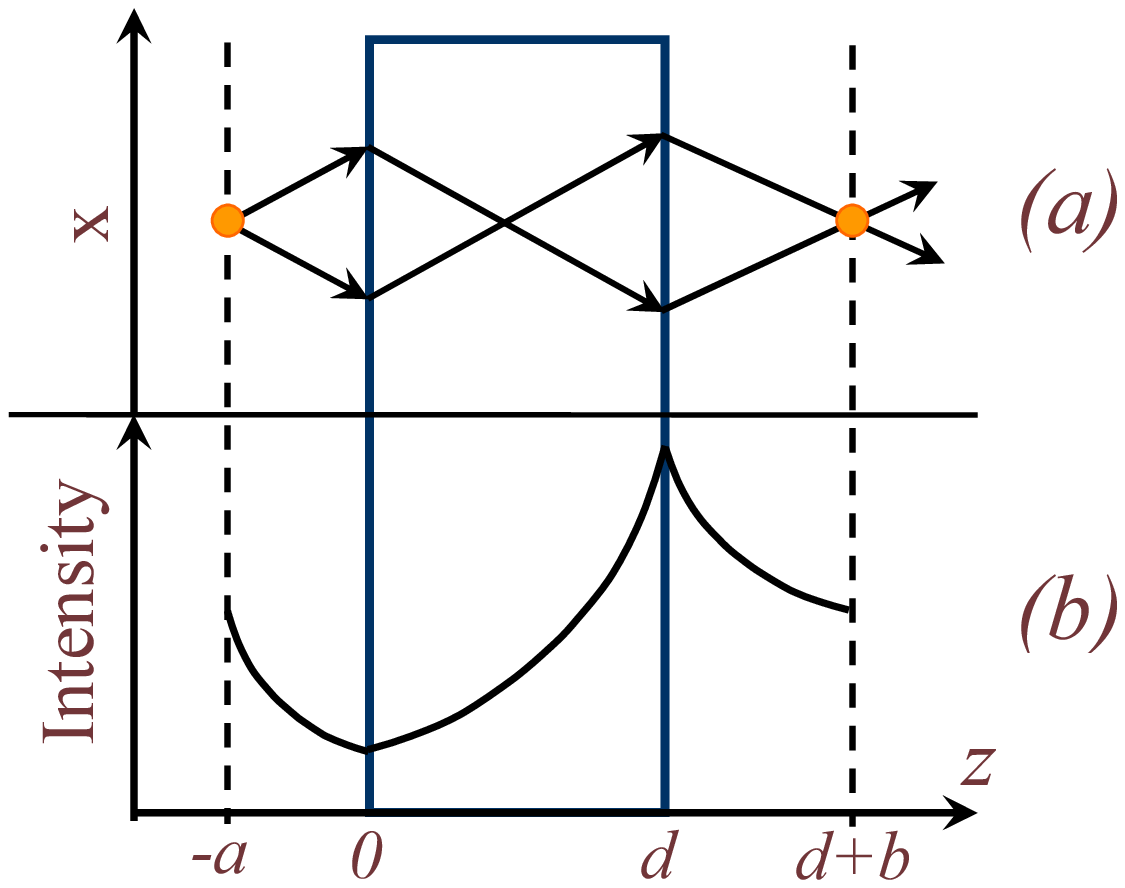}{fig1}{(color online) (a) Ray focusing and (b)
schematic reconstruction of evanescent waves by a left-handed
lens.}

Thus, a slab of the left-handed metamaterial can be used for a
sub-wavelength imaging because it amplifies all evanescent modes (near field) inside the slab, and therefore allows to preserve the
information about the source structure with the resolution better than the
radiation wavelength. However, to satisfy the conditions for such
a perfect lens to operate, the distance between a source and the
slab surface, $a$, and the distance
between the second surface of the slab and the image point, $b$,
should be connected with the slab thickness $d$ by the
relation~\cite{ref1_anis} (see Fig.~\rpict{fig1}),
\be
\label{eq1_anis}
a + b = d.
\ee
The relation (\ref{eq1_anis}) means that it is impossible to
create an image at the distances larger than the slab thickness,
and this is one of the serious limitations for applicability of a left-handed perfect lens.

In this Letter, we introduce a concept of the birefringent
non-reflecting left-handed metamaterials and {\em birefringent
perfect lenses}. In particular, we show that, in contrast to the
conventional perfect lens condition $\epsilon=\mu=-1$, the
birefringent left-handed lens can focus either TE or TM polarized
waves or both of them, with {\em a varying distance} between the
TE and TM images; and this property allows to expand dramatically
the applicability limits of the perfect lenses. In addition, we
show that such a birefringent lens is free from the limitations
imposed by the condition~(\ref{eq1_anis}), and we also discuss some
other applications of the birefringent left-handed metamaterials
for the beam polarization splitting and sub-wavelength beam
imaging.

We consider a linear medium described by the following
tensors of dielectric permittivity $\hat{\epsilon}$ and magnetic
permeability $\hat{\mu}$, which in the main axes of the crystal have the form
\be \label{eq2_anis}
\hat{\epsilon}= \left( \begin{array}{ccc} A & 0 & 0 \\
0 & B & 0 \\ 0 & 0 & A^{-1} \end{array} \right), \;\;\;
\hat{\mu}= \left( \begin{array}{ccc} B & 0 & 0 \\
0 & A & 0 \\ 0 & 0 & B^{-1} \end{array} \right) \ee
where $A$ and $B$ are generally arbitrary complex functions of the
frequency. We substitute the expressions (\ref{eq2_anis}) into
Maxwell's equations and obtain the equations for the transverse spatial harmonics of the monochromatic [$\sim \exp (i \omega t - i k_x x)$] electromagnetic
waves, for the case of (i) {\em TM polarization}, when ${\bf E} = (E_x,
0, E_z)$ and ${\bf H} = (0, H_y, 0)$:
\be \label{eq3_anis} \frac{d^2H_y}{dz^2} + A^2(k_0^2 -k_x^2)
H_y=0, \ee
\be \label{eq4_anis} E_x = - \frac{1}{ik_0A} \frac{dH_y}{dz},
\;\;\; E_z = -\frac{k_x}{k_0} A H_y, \ee and for the case of (ii)
{\em TE polarization} when ${\bf E} = (0, E_y, 0)$ and  ${\bf H} = (H_x,
0, H_z)$:
\be \label{eq5_anis} \frac{d^2E_y}{dz^2} + B^2(k_0^2 -k_x^2)
E_y=0,
\ee
\be \label{eq6_anis} H_x = - \frac{1}{ik_0B} \frac{dE_y}{dz},
\;\;\; H_z = \frac{k_x}{k_0} B E_y, \ee where $k_0 =\omega/c$ is
the wave number in vacuum, $k_x$ is the wave vector component
along the $x-$axes, and $c$ is the speed of light. It is easy to
verify that the wave impedance of this birefringent medium matches
exactly the impedance of vacuum, for both the polarizations with
any transverse wavenumbers $k_x$, and for {\em arbitrary}
(including complex) values of $A$ and $B$. Therefore, the medium described by the tensors (\ref{eq2_anis}) is ideally impedance-matched with vacuum being reflectionless~\cite{ref2_anis}. Such a birefringent medium was suggested as a perfectly matched layer in the finite-difference
time-domain simulations~\cite{Sacks:1995-1460:ITAP}. In a general
case, when the vacuum is substituted by a medium with some
$\epsilon_s$ and $\mu_s$, the impedance matching conditions would
require some modification of Eq.~(\ref{eq2_anis}), namely
$\hat{\epsilon} \rightarrow \epsilon_s \hat{\epsilon}$, and
$\hat{\mu} \rightarrow \mu_s \hat{\mu}$. Below we consider
$\epsilon_s =\mu_s =1$ without loss of generality.

\pict{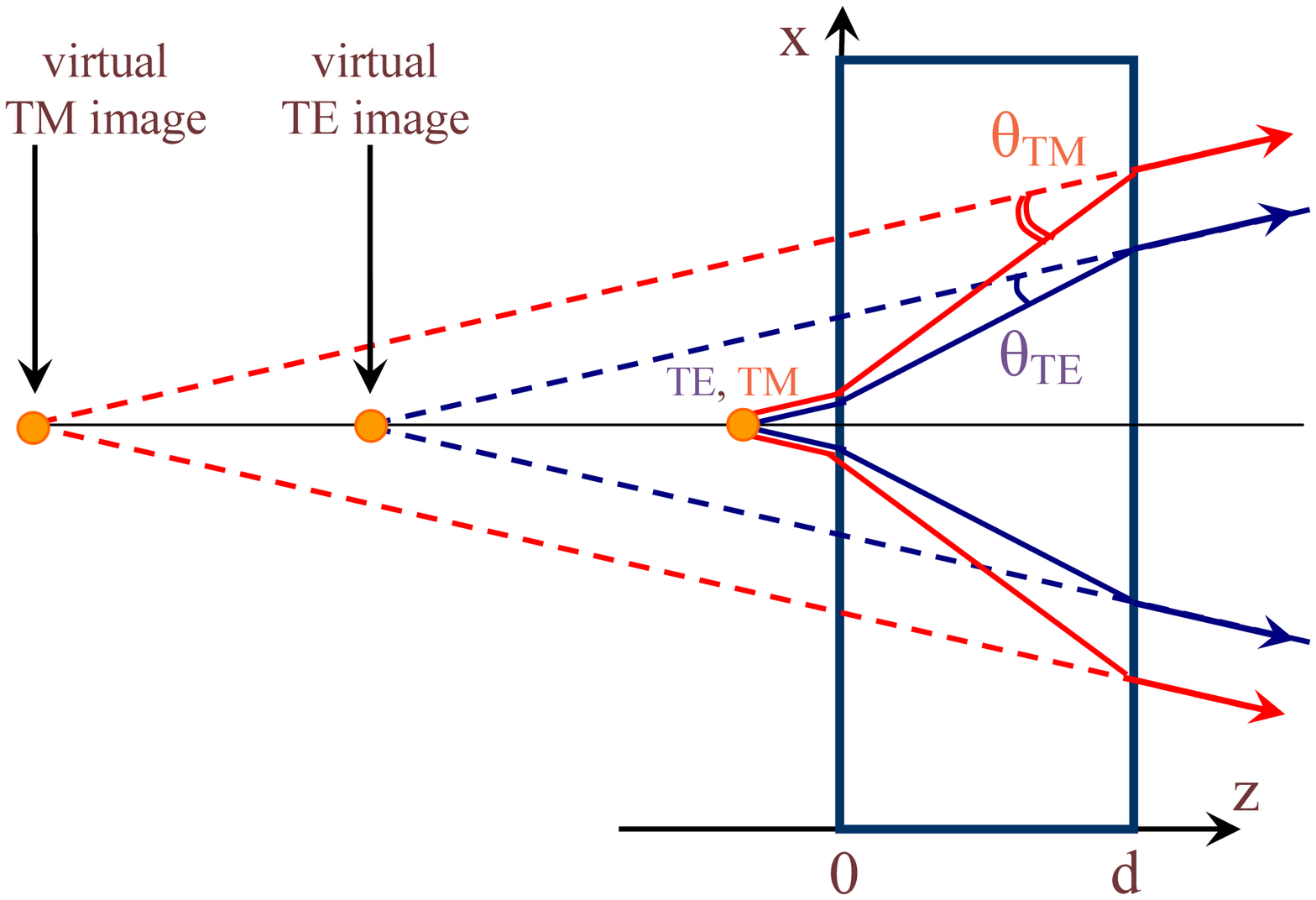}{fig2}{(color online) Ray diagram showing the
creation of two separate TE and TM polarized virtual images of a
source; $\theta_{TE}$ and $\theta_{TM}$ are the angles between the
group and phase velocities of the TE and TM waves inside the
slab.}

We consider a slab of the metamaterial with the dielectric and
magnetic properties characterized by the tensors (\ref{eq2_anis}).
The slab is surrounded by vacuum, and it has the thickness $d$ ($0
\leq z \leq d$). We assume that a point source is located at the
distance $z=-a$ from the nearest surface of the slab, as shown in
Fig.~\rpict{fig2}, the source generates both TE and TM polarized
waves, and it is described by the corresponding distribution of
the electric field $E_y(x, z =-a)$, for the TE polarization, or
the magnetic field $H_y(x, z=-a)$, for the TM polarization, in the
plane $z=-a$. We denote the spatial spectra of the these fields as
$\alpha_e(k_x)$ and $\alpha_m(k_x)$, respectively. Using
Eqs.~(\ref{eq3_anis}) to (\ref{eq6_anis}) for describing the
electromagnetic field in the slab, and satisfying the boundary
conditions for the tangential components of the fields, we obtain
the general expressions for the spatial harmonics of the fields
behind the slab, i.e. for $z> d$,
\be \label{eq8_anis} H_y(z,k_x) = \alpha_m(k_x) \exp \left\{ -i
\sqrt{k_0^2 -k_x^2} (a + Ad + z^{\prime})\right\}, \ee
for the TM polarized waves, and
\be \label{eq9_anis} E_y(z,k_x) = \alpha_e(k_x) \exp \left\{ -i
\sqrt{k_0^2 -k_x^2} (a + Bd + z^{\prime})\right\}, \ee
for the TE polarized waves, where $z^{\prime}=z-d$. Thus, for real
$A$ and $B$ Eqs.~(\ref{eq8_anis}),~(\ref{eq9_anis}) reproduce the
field structure of the source in the region $z^{\prime}>0$ shifted from the source position by the distance $(A-1)d$ (for TM waves) or $(B-1)d$ (for TE waves). A typical ray diagram for this case is shown in Fig.~\rpict{fig2} for $A>1$, $B>1$ and $A>B$, where we show the position of the source for both the polarizations, as well as spatially separated
virtual images created by the lens. In general, for $A \neq B$,
the virtual images of the TE and TM sources are shifted relative
to each other. For $0 < A, B < 1$, the virtual images can be located either
between the slab and the source or inside the metamaterial slab.

More interesting cases of the medium (\ref{eq2_anis}) correspond
to {\em negative} values of $A$ or/and $B$. When $A<0$ and $B>0$,
negative refraction occurs for the TM polarized waves only, whereas the
TE polarized waves refract normally, see Fig.~\rpict{fig3}. For
$A>0$ and $B<0$, the opposite effect occurs, i.e. negative
refraction is possible for the TE polarized waves only. This
property can be used for the polarization-sensitive beam
separation. Figure~\rpict{fig3} shows an example of this
separation for the slab with $A=-4$ and $B=+2$. A two-dimensional
beam propagates at the angle of incidence 30$^o$ and is
refracted. Initially, the beam is composed of two polarizations with the same
partial intensities. When the beam is refracted at the surface,
the TM polarized wave undergoes negative refraction and it becomes
separated from the normally refracted TE beam.

\pict{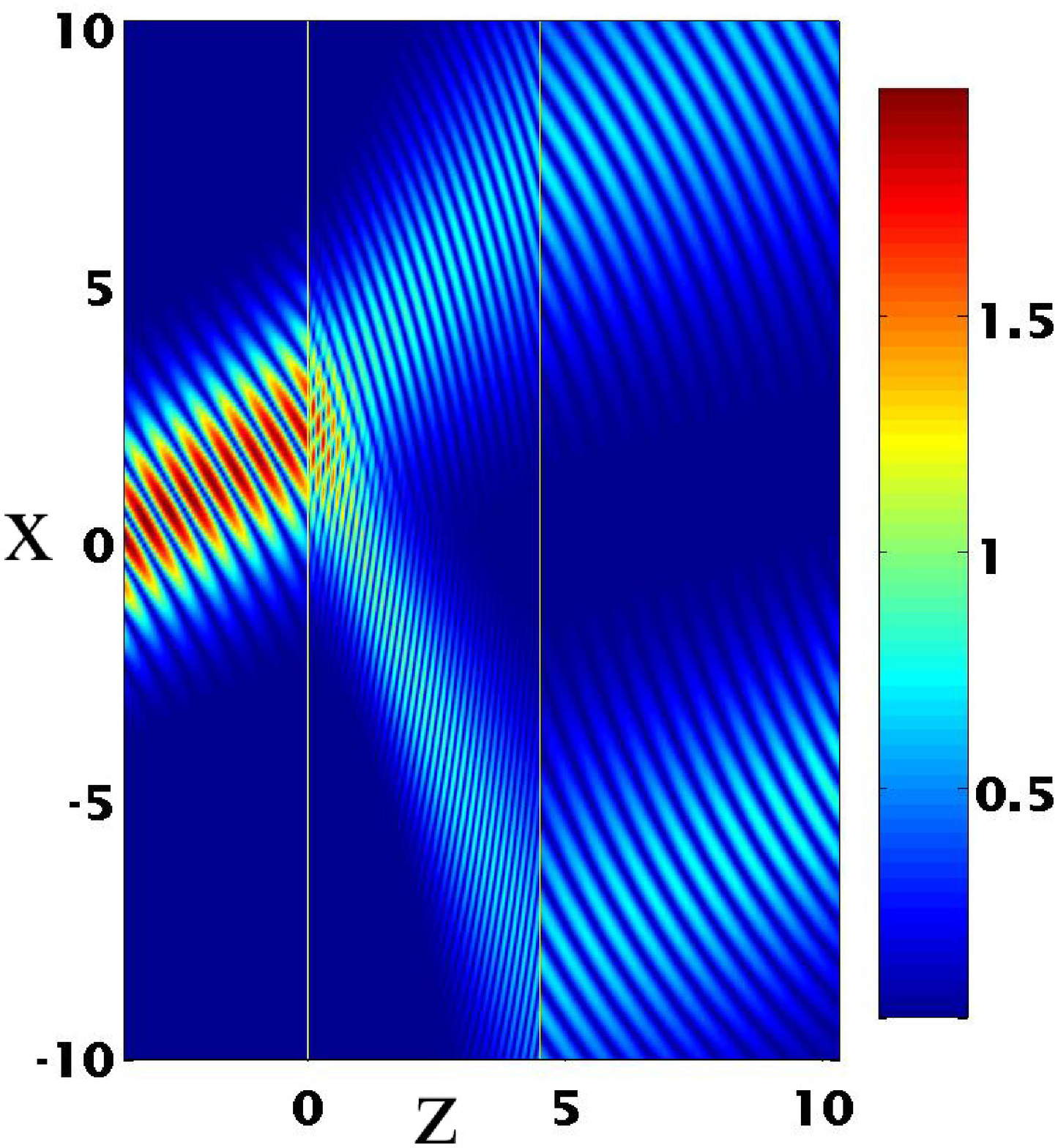}{fig3}{(color online)  Beam transmission through a
slab of the birefringent metamaterial ($A=-4$, $B=+2$, $d=5$). The
TM polarized component experiences negative refraction, while the
TE polarized component refracts normally. Coordinates are
normalized to the free-space wavelength.}

Another specific feature of the birefringent lenses is a
possibility to form {\em two separate perfect images} for the TE and TM
polarized waves. This property follows from the results
(\ref{eq8_anis}) and (\ref{eq9_anis}). In particular, for $A<0$
and $B>0$, the transverse spatial spectrum of the TM polarized
field in the plane $z_m^{\prime} = |A|d -a$ coincides with the
spectrum of the source,
\be
\label{eq10_anis}
H_y(z_m^{\prime}, k_x) = \alpha_m(k_x),
\ee
while the TE polarized component of the beam is positively refracted. In the case $A>0$ and $B<0$, the image is created by the TE polarized waves at
$z_e^{\prime} = |B|d -a$,
\be
\label{eq11_anis}
E_y(z_e^{\prime}, k_x) = \alpha_e(k_x),
\ee
whereas the TM polarized waves experience positive refraction, and they do not create an image. Thus, in the case of the birefringent lens additional parameters appear, which mitigate the strict limitations for the isotropic lens imposed by Eq.~(\ref{eq1_anis}). As a result, the source and the image can be located further away from the slab. More importantly, when both $A$ and
$B$ are negative and $A \neq B$, both TE and TM images appear, and
they are separated by the distance
\be \label{eq12_anis} h = |z_e^{\prime}-z_m^{\prime}| =
\left|(|B|-|A|)d\right|,
\ee
which, in the absence of dissipative losses, can be arbitrary
large. This allows novel possibilities for sub-wavelength
resolution, diagnostics, and microscopy.

\pict{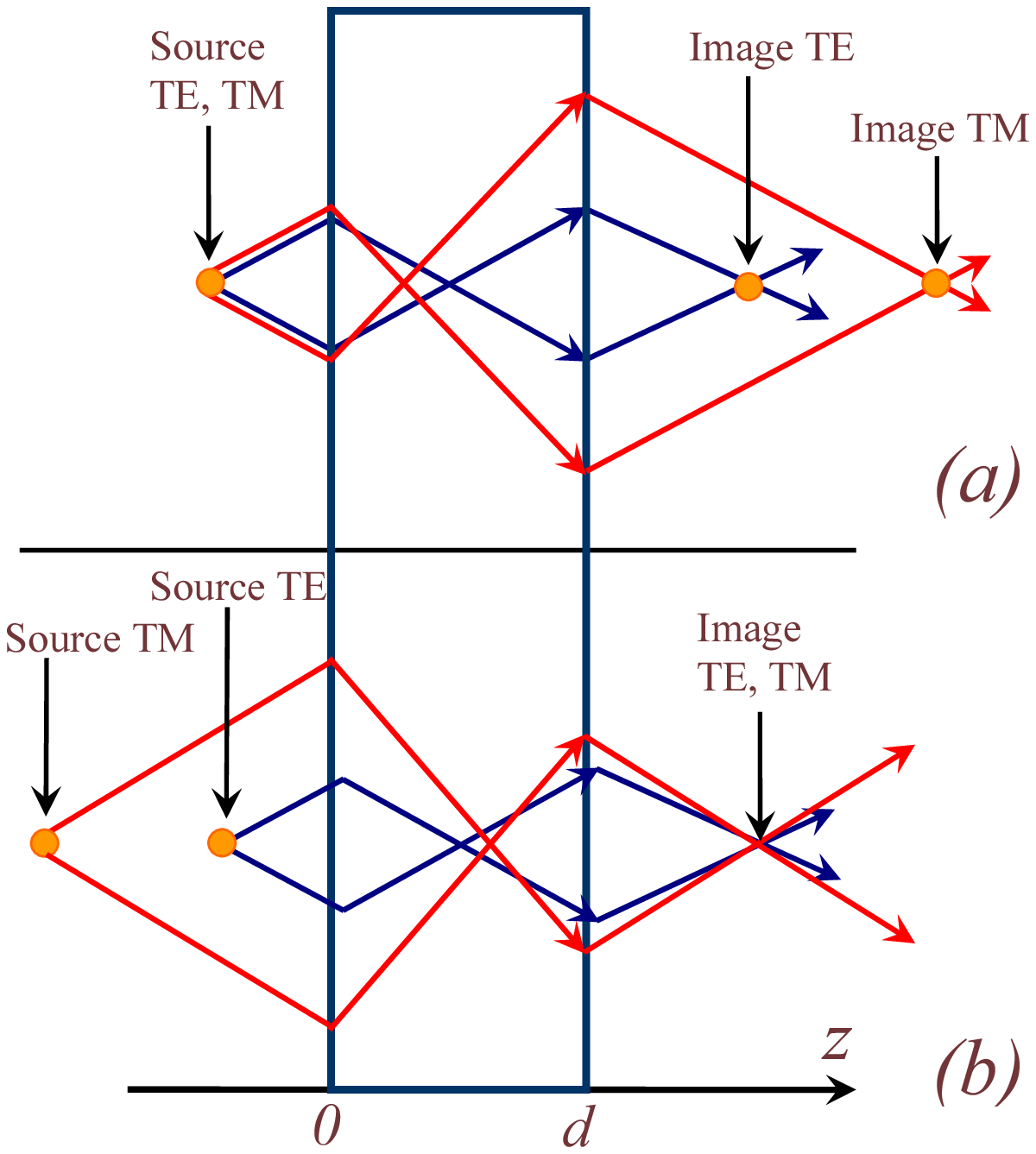}{fig4}{(color online) Ray diagrams of the
birefringent left-handed lens imaging. (a) A single source imaging
by a metamaterial slab characterized by negative $A \ne B$. TE and
TM images are separated by the distance $h$ defined by
Eq.~(\ref{eq12_anis}). (b) Separated sources with two different
polarizations can create the images in the same plane, provided
the sources are separated by the distance~(\ref{eq12_anis}).}

In the case $|A| =|B|$, the TE and TM images coincide and in a
particular case $A=-1$ and $B=-1$ we recover the results for the {\em
isotropic perfect lens} discussed by Pendry~\cite{ref1_anis} and
Veselago~\cite{ref3_anis}. In general, the basic physics for
operating the birefringent perfect lenses is similar to the
isotropic case, and it is defined by two major factors: (i)
negative refraction, and (ii) amplification of evanescent waves.
Figure~\rpict{fig1}(b) shows schematically the structure of the
evanescent waves in the slab for the case of Pendry's lens, which
is similar for both isotropic and birefringent left-handed media.
Figures~\rpict{fig4}(a,b) show schematically the ray diagram in
two special cases, when a single source generates both TE and TM
polarized waves [see Fig. \rpict{fig4}(a)] creating two separate
images, and when the TE and TM sources are separated and they
create a combined image [see Fig.~\rpict{fig4}(b)].

A possibility of the sub-wavelength resolution of a pair of subwavelength sources by using the birefringent left-handed lens has also been verified numerically, and some examples are presented in Fig.~\rpict{fig6} for the case
of a lossy medium when $\hat{\epsilon}_{l} = \hat{\epsilon}-i\delta_{ik}\times 10^{-8}$ and $\hat{\mu}_{l} =\hat{\mu} - i\delta_{ik}\times 10^{-8}$, where $\delta_{ik}=1$ for $i=k$ and it is $0$ otherwise. The mixed-polarized source consists of two beams of the width $\lambda/5$, separated by the distance $2 \lambda/5$, where $\lambda$ is the free-space wavelength, metamaterial parameters are $A=-2.5$ and $B=-1.5$, and the slab thickness is $\lambda/2$. A difference in the resolution for the TE and TM polarized waves for $A \neq B$ is explained by different effective
optical thickness of the slab for two polarizations. An increase
of the losses decreases the resolution abilities of the lens dramatically.
Figures~\rpict{fig5}(a,b) show the spatial distribution of the
magnetic field, ${\cal R}e \left( H_y(x,z) \right)$, for the TM
polarization, and spatial distribution of electric field ${\cal
R}e \left( E_y(x,z) \right)$, for the TE polarization,
respectively. 

\pict{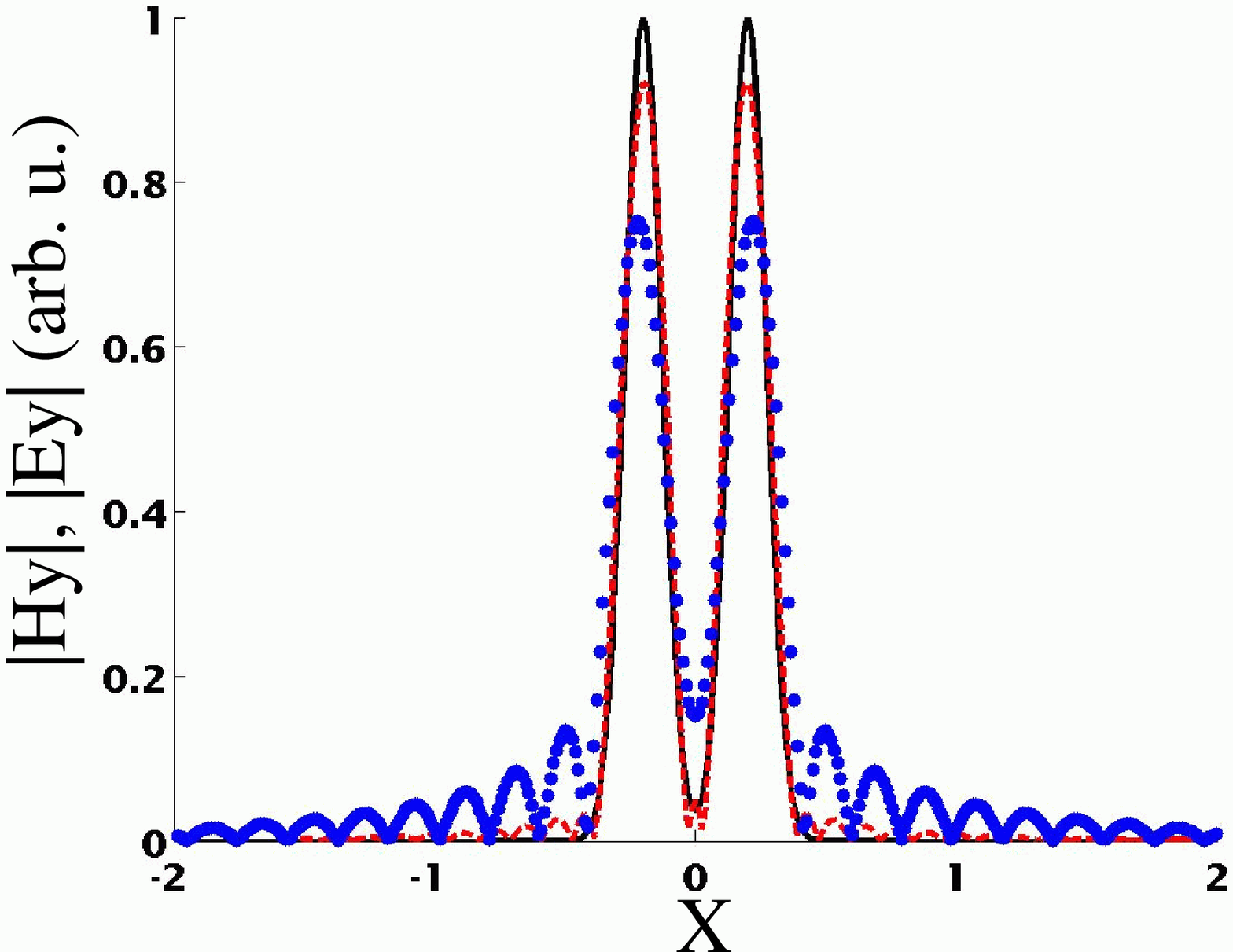}{fig6}{(color online) TE and TM fields of the
source (solid), electric field of the TE image (dashed) and magnetic field of the  TM image (dotted). The coordinate is normalized to the free-space wavelength.}

\pict{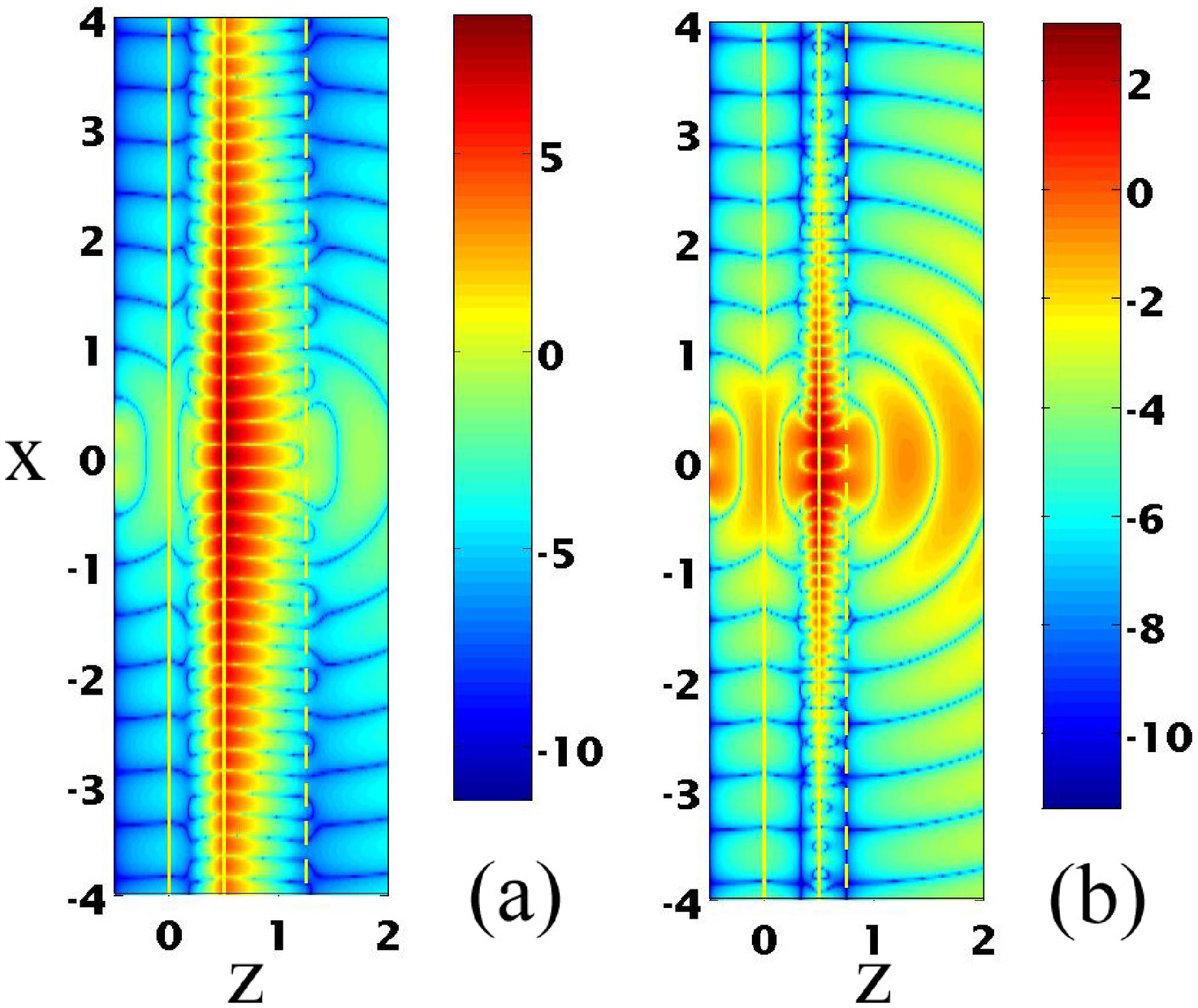}{fig5}{(color online) Spatial distribution of the absolute values of (a) magnetic and (b) electric fields (logarithmic scale) generated by the two subwavelength sources of the wifth $\lambda/5$ separated by $2 \lambda/5$. Parameters are $A=-2.5$ and $B=-1.5$, and the slab thickness is $\lambda/2$. Solid lines mark the metamaterial slab, dashed lines show the image planes.
Coordinates are normalized to the free-space wavelength.}

Different examples presented above clearly demonstrate that the
birefringent left-handed metamaterials and birefringent perfect
lenses are novel objects with many unusual properties and, more
importantly, they may demonstrate much broader spectrum of potential applications, in comparison with the isotropic metamaterials and perfect
lenses~\cite{ref1_anis,ref3_anis}. Although the birefringent perfect
lenses are not yet realized in experiment, we believe that the ideas and results presented here are quite realistic and will initiate strong efforts in creating the composite metamaterials with {\em substantially birefringent}
properties, including those that satisfy the specific conditions
for the tensor components of Eq.~(\ref{eq2_anis}). This would
require a new thinking in applying the traditional
approaches~\cite{ref4_anis,ref5_anis} where the fabrication of
isotropic metamaterials was made the main target. Such an anisotropy can
be achieved by using more complicated elementary cells made of
wires and split-ring resonators, instead of the traditional
symmetric cubic lattice~\cite{ref5_anis}, in order to engineer
both the electric and magnetic response in three different directions. We also note that in order to realize a birefringent lens, which is able to create an image of a three-dimensional source (compared to the two-dimensional case considered above), one should take the metamaterial with $A=B$, and it can simplify the design of the composite. Such lens creates the image of both wave polarizations in the same plane.

In conclusion, we have introduced a novel type of birefringent left-handed
media which possess a number of unique properties, including reflectionless scattering independent on a type of incoming monochromatic waves, focusing and negative refraction that occur under different conditions for the TE and TM polarized waves or simultaneously with two spatially separated TE and TM images. We believe our results suggest novel directions in the study of the intriguing properties of metamaterials and their fabrications.

The authors acknowledge a support of the Australian Research Council, and thank A.I. Smirnov for a useful reference. AAZ and REN thank I.G. Kondrat'ev for discussions. AAZ and NAZ acknowledge a warm hospitality of the Nonlinear Physics Centre in Canberra. AAZ acknowledges a support from RFBR (grant N05-02-16357).

\end{sloppy}


\begin{thebibliography}{}

\bibitem{ref1_anis} J.B. Pendry, Phys. Rev. Lett. {\bf 85}, 3966 (2000).

\bibitem{ref3_anis} V.G. Veselago, Usp. Fiz. Nauk {\bf 92}, 517 (1967) [Sov. Phys. Uspekhi {\bf 10}, 569 (1968)].

\bibitem{review} D.R. Smith, J.B. Pendry, and M.C.K. Wiltshire,
Science {\bf 305}, 788 (2004).

\bibitem{critics} See, e.g., G.W. 't Hooft, Phys. Rev. Lett. {\bf 87}, 249701 (2001); N. Garcia and M. Nieto-Vesperinas, Phys. Rev. Lett. {\bf 88},
207403 (2002).

\bibitem{more} J.T. Shen and P.M. Platzman, Appl. Phys. Lett. {\bf 80}, 3286
(2002); D.R. Smith, D. Schurig, M. Rosenbluth, S. Schultz, S.A.
Ramakrishna, and J.B. Pendry, Appl. Phys. Lett. {\bf 82}, 1506
(2003).

\bibitem{numerics} P. Kolinko and D.R. Smith, Opt. Express {\bf 11}, 640
(2003); S.A. Cummer, Appl. Phys. Lett. {\bf 82}, 1503 (2003).

\bibitem{george} A. Grbic and G.V. Eleftheriades, Phys. Rev. Lett. {\bf 92}, 117403
(2004).

\bibitem{ref2_anis} Reflectionless of this kind of media but with
$A=B>0$ has been mentioned earlier in S.P. Efimov, Izv. VUZov
Radiofizika {\bf 21}, 1318 (1978) [Radiophys. Quantum Electron.
{\bf 21}, 1318 (1978)].

\bibitem{Sacks:1995-1460:ITAP} Z. Sacks, D. Kingsland, R. Lee, and J. Lee, IEEE Trans. Antennas and Propagation  {\bf 43}, 1460 (1995).

\bibitem{ref4_anis} R.A. Shelby, D.R. Smith, and S. Schultz, Science {\bf 292}, 77 (2001).

\bibitem{ref5_anis} C.G. Parazzoli, R.B. Greegor, K. Li, B.E.C. Koltenbah, and M. Tanielian, Phys. Rev. Lett. {\bf 90}, 107401 (2003).


\end{thebibliography}
\end{document}